\documentclass[structabstract]{aa}  
\usepackage{graphicx}
\usepackage{txfonts}
\usepackage{natbib}
\usepackage{lscape}

\begin{document}
   \title{Ultraluminous X-ray sources out to z$\sim0.3$ in the COSMOS field.}


   \author{V.~Mainieri\inst{1},
          C.~Vignali\inst{2,3},
          A.~Merloni\inst{4,5},
          F.~Civano\inst{6},
          S.~Puccetti\inst{7},
	  M.~Brusa\inst{5},
	  R.~Gilli\inst{3},
	  M.~Bolzonella\inst{3},
	  A.~Comastri\inst{3},
          G.~Zamorani\inst{3},
	  M.~Aller\inst{9},
	  M.~Carollo\inst{9},
	  C.~Scarlata\inst{9,10},
	  M.~Elvis\inst{6},
	  T.L.~Aldcroft\inst{6},
	  N.~Cappelluti\inst{5},
          G.~Fabbiano\inst{6}, 
	  A.~Finoguenov\inst{5},
	  F.~Fiore\inst{8},
	  A.~Fruscione\inst{6},
	  A.M.~Koekemoer\inst{11},
	  T.~Contini\inst{12},
	  J.-P.~Kneib\inst{13},
	  O.~Le F\`{e}vre\inst{13},
	  S.~Lilly\inst{9},
	  A.~Renzini\inst{14},
	  M.~Scodeggio\inst{15},
	  S.~Bardelli\inst{3},
	  A.~Bongiorno\inst{5},
	  K.~Caputi\inst{9},
	  G.~Coppa\inst{3},
	  O.~Cucciati\inst{16},
	  S.~de la Torre\inst{13},
	  L.~de Ravel\inst{13},
	  P.~Franzetti\inst{15},
	  B.~Garilli\inst{15},
	  A.~Iovino\inst{16},
	  P.~Kampczyk\inst{9},
	  C.~Knobel\inst{9},
	  K.~Kova\v{c}\inst{9},
	  F.~Lamareille\inst{12},
	  J.-F.~Le Borgne\inst{12},
	  V.~Le Brun\inst{13},
	  C.~Maier\inst{9},
	  M.~Mignoli\inst{3},
	  R.~Pello\inst{12},
	  Y.~Peng\inst{9},
	  E.~Perez Montero\inst{12},
	  E.~Ricciardelli\inst{17},
	  J.~D.~Silverman\inst{9},
	  M.~Tanaka\inst{1},
	  L.~Tasca\inst{13},
	  L.~Tresse\inst{13},
	  D.~Vergani\inst{3},
	  E.~Zucca\inst{3},
          P.~Capak\inst{19},
	  O.~Ilbert\inst{13},
	  C.~Impey\inst{18},
	  M.~Salvato\inst{19},
	  N.~Scoville\inst{19},
          Y.~Taniguchi\inst{20},
	  J.~Trump\inst{18}
          }

   \institute{ESO, Karl-Schwarschild-Strasse 2, D--85748 Garching 
     bei M\"unchen, Germany
	\and 
             Dipartimento di Astronomia, Universit\`a degli Studi
  di Bologna, Via Ranzani 1, I--40127 Bologna, Italy
	\and
		INAF$-$Osservatorio Astronomico di Bologna, Via
  Ranzani 1, I--40127 Bologna, Italy
	\and
		Excellence Cluster Universe, TUM, Boltzmannstr. 2,
		D-85748 Garching  bei M\"unchen, Germany
	\and
		Max-Planck-Institute f\"ur Extraterrestrische Physik,
  Postfach 1312, 85741, Garching bei M\"unchen, Germany
	\and
	Harvard-Smithsonian Center for Astrophysics, 60
  	Garden St., Cambridge, MA 02138 USA
	\and
	ASI Science Data Center, via Galileo Galilei, 00044
  Frascati Italy
	\and
	INAF$-$Osservatorio astronomico di Roma, Via Frascati
  33, 00040 Monteporzio Catone, Italy
	\and
	Institute of Astronomy, Swiss Federal Institute of Technology (ETH H\"onggerberg), CH-8093, Z\"urich, Switzerland
	\and
        Spitzer Science Center, Pasadena, CA, 91125
        \and
	Space Telescope Science Institute, Baltimore, Maryland 21218, USA
	\and
	Laboratoire d'Astrophysique de Toulouse-Tarbes, Universit\'{e} de Toulouse, CNRS, 14 avenue Edouard Belin, F-31400 Toulouse, France
	\and
	Laboratoire d'Astrophysique de Marseille, Marseille, France
	\and
	 INAF$-$Osservatorio astronomico di Padova, Vicolo Dell'Osservatorio 5, I-35122 Padova, Italy
	\and
	INAF - IASF Milano, Milan, Italy
	\and
	INAF Osservatorio Astronomico di Brera, Milan, Italy
	\and 
	Dipartimento di Astronomia, Universita di Padova, Padova, Italy
	\and
	Steward Observatory, University of Arizona, 933 North Cherry Avenue, Tucson, AZ 85721
	\and
	California Institute of Technology, MC 105-24, 1200
  East California Boulevard, Pasadena, CA 91125 UA
        \and
        Research Center for Space and Cosmic Evolution,
  Ehime University, Bunkyo-cho 2-5, Matsuyama 790-8577, Japan
             }

   \date{Received ...; accepted ...}

 
  \abstract
   {Using Chandra observations we have identified a sample of seven off-nuclear X-ray sources, in the redshift range z=0.072-0.283, located within optically bright galaxies in the COSMOS Survey. All of them, if associated with their closest bright galaxy, would have  L$[0.5-7$ keV$] >10^{39}$ erg s$^{-1}$ and therefore can be classified as ultraluminous X-ray sources (ULXs). }
   {Using the multi-wavelength coverage available in the COSMOS field, we study the properties of the host galaxies of these ULXs. In detail, we derived their star formation rate from H$\alpha$ measurements and their
stellar masses using SED fitting techniques with the aim to compute the
probability to have an off-nuclear source based on the host galaxy
properties. We divide the host galaxies in different morphological classes using the available ACS/HST imaging.}
{We selected off-nuclear candidates
 with the following criteria: 1) the distance between the X-ray and the optical centroid has to be larger than 0.9\arcsec, larger than 1.8 times the radius
 of the Chandra positional error circle and smaller than the Petrosian radius of the host galaxy; 2) the optical counterpart is a bright galaxy (R$_{\rm AB}<$22); 3) the redshift of the counterpart is lower than z$=0.3$; 4) the source has been observed in at least one Chandra pointing at an off-axis angle smaller than 5\arcmin; 5) the X-ray positional error is smaller than 0.8\arcsec. We verified each candidate super-imposing the X-ray contours on the optical/IR images. We expect less than one misidentified AGN due to astrometric errors and on average 1.3 serendipitous background source matches.}
   {We find that our ULXs candidates are located in regions of the SFR versus M$_\star$ plane where one or more off-nuclear detectable sources are expected. From a morphological analysis of the ACS imaging and the use of rest-frame colours, we find that our ULXs are hosted both in late and early type galaxies. Finally, we find that the fraction of galaxies hosting a ULX ranges from $\approx 0.5\%$ to $\approx0.2 \%$ going from L$_{0.5-2~keV}=3\times 10^{39}$ erg s$^{-1}$ to L$_{0.5-2~keV}=  2 \times 10^{40}$ erg s$^{-1}$. }
   {}

  \authorrunning{V. Mainieri}
  \titlerunning{ULXs in COSMOS}
   \maketitle
%



   \begin{figure}
   \centering
   \includegraphics[width=8cm]{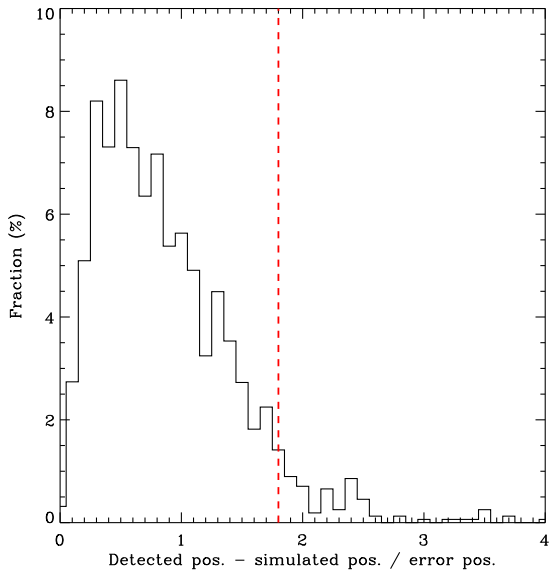}
      \caption{The distribution of the difference between the detected X-ray positions and the input positions in units of the X-ray positional error. 
              }
         \label{pos-acc}
   \end{figure}

\section{Introduction}

An intriguing class of X-ray objects are the so called ultraluminous
X-ray sources (ULXs). Here an ULX is defined as an X-ray source in an
extra-nuclear region of a galaxy with an observed luminosity in excess
of $10^{39}$ erg s$^{-1}$ in the 0.5-7 keV band. Such X-ray
luminosities are higher than expected for spherical Eddington-limited
accretion onto a $\sim 10 M_\odot$ black hole. ULXs were known already
from studies with Einstein, ROSAT, and ASCA (e.g. \citealt{fabbiano89,
  colbert02, makishima00}), but it was after the advent of Chandra
with its combination of high angular resolution and moderate spectral
resolution that has been possible to make significant progresses in
their study (e.g. \citealt{roberts04,swartz04}). There is a wide
debate in the literature on the nature of these sources. ULXs may be
powered by accretion onto stellar-mass black holes assuming that there
is relativistic beaming (e.g. \citealt{kording02}), or radiative
anisotropy (e.g. \citealt{king02}), or they may be associated with
super-Eddington discs (e.g. \citealt{begelman02}). It has been also
suggested that ULXs represent a new class of intermediate-mass
($10^2-10^5$ M$_\odot$) black holes
(e.g. \citealt{colbert99,miller04}). These intermediate-mass black
holes may be fed by Roche lobe overflow from a tidal captured stellar
companion that is not destroyed by tidal heating
(\citealt{hopman04}). Off-nuclear AGN activity could be also a
signature of a recoiling massive black hole: a massive black hole
binary coalesces and gives origin to gravitational waves which can
give a kick to the center of mass of the system. If the recoiling
black hole retains the inner parts of its accretion disk, we could see
its luminous phase as an off-nuclear AGN (see \citealt{volonteri08}
and references therein). Finally, ULXs could also be the
high-luminosity extension of supernovae
(e.g. \citealt{swartz04}).\\ Many of the previous studies based on
Chandra data are focused on local galaxies, where the Chandra angular
resolution allows to detect several off-nuclear sources in one single
galaxy. In this paper, we select a sample of ULXs from the Chandra
survey in the COSMOS field. We have here the advantage to combine deep
X-ray observation with a wealth of multiwavelength ancillary data that
we will use to put constraints on the nature of these sources and on
the properties of their host galaxies. The redshift range that we are
covering is up to $z\simeq0.3$. A study on off-nuclear sources in a
similar redshift range was performed by \citet{lehmer06} on the
Chandra Deep Fields (CDFs).\\ We quote in this paper magnitudes in the
AB system and we assume a cosmology with H$_0 = 70$ km s$^{-1}$
Mpc$^{-1}$, $\Omega _{\rm M} = 0.3$ and $\Omega _\Lambda = 0.7$.


   \begin{figure}
   \centering
   \includegraphics[width=8cm]{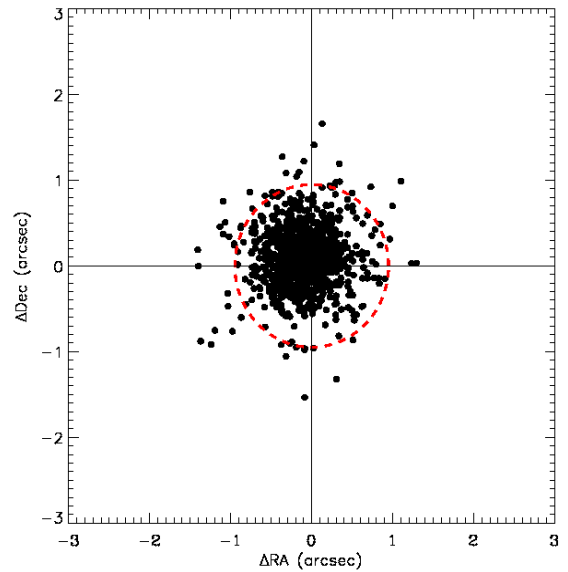}
      \caption{X-ray to optical offsets in arcsec for X-ray sources with a secure identification (\citealt{civano09}) and with X-ray positional error smaller than 0.8\arcsec. The circle of 0.9\arcsec radius encompasses $95\%$ of the X-ray sources.
              }
         \label{ast-acc}
   \end{figure}

   \begin{figure*}
   \centering
   \includegraphics[width=16cm]{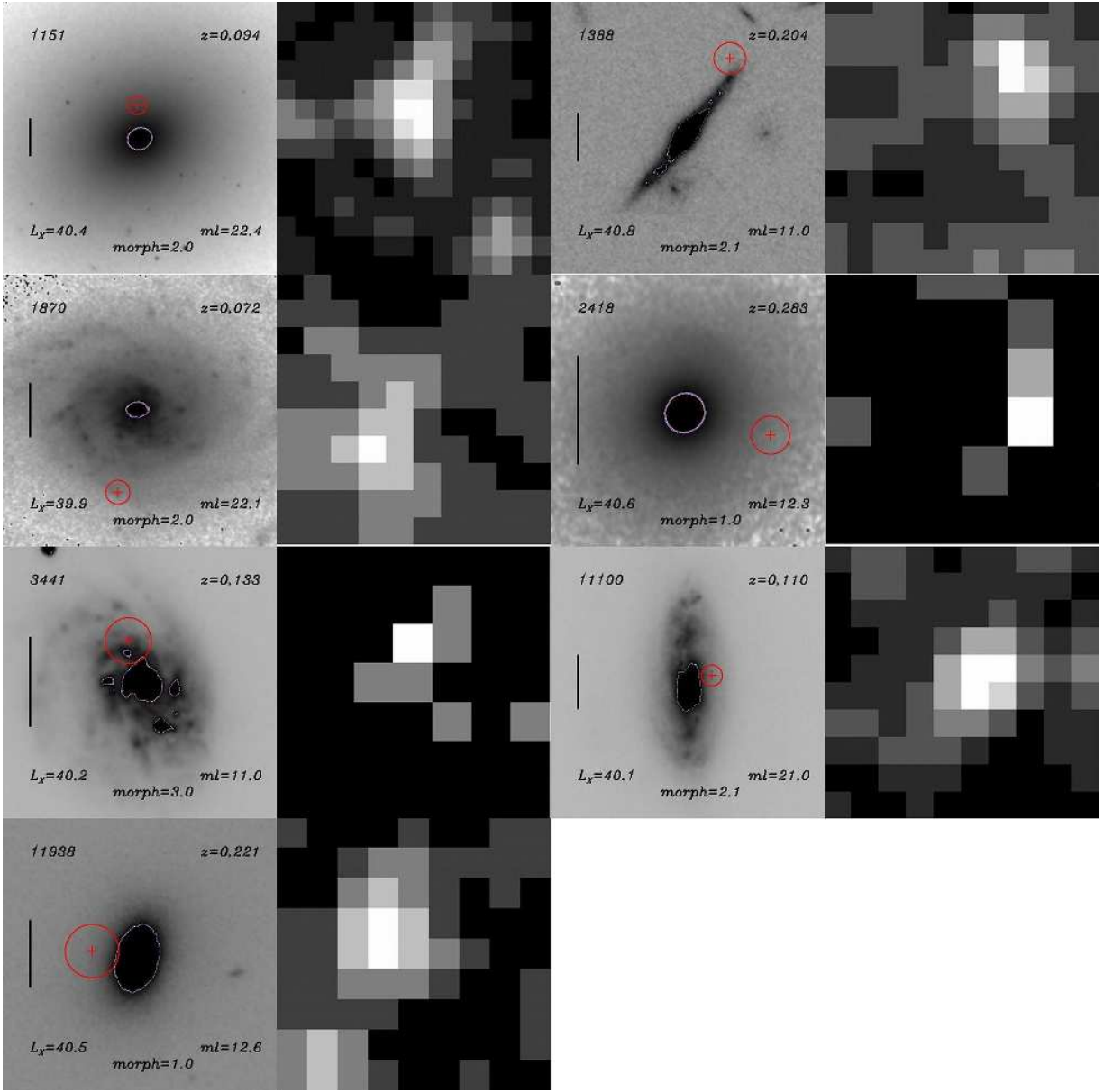}
  \caption{Cutouts in the HST/ACS F814W band (\citealt{koekemoer07}) of the seven X-ray off-nuclear sources in the C-COSMOS field. The red cross indicates the position of the X-ray centroid and the red circle the X-ray positional error (\citealt{elvis09}). We provide for each object: the Chandra ID (top-left), the redshift (top-right), the logarithm of the X-ray luminosity in the [0.5-7] keV band (bottom-left), the maximum likelihood ratio for the X-ray detection (bottom-right), the morphological classification of the host galaxy (bottom-middle; see Section \ref{section:classification}). The images have different sizes for display purposes; the vertical bar in each cutout corresponds to 2\arcsec~. On the right of each ACS cutout, there is the corresponding Chandra [0.5-7 keV] image.}
              \label{mosaic1}%
    \end{figure*}
   \begin{figure*}
   \centering
   \includegraphics[width=10cm]{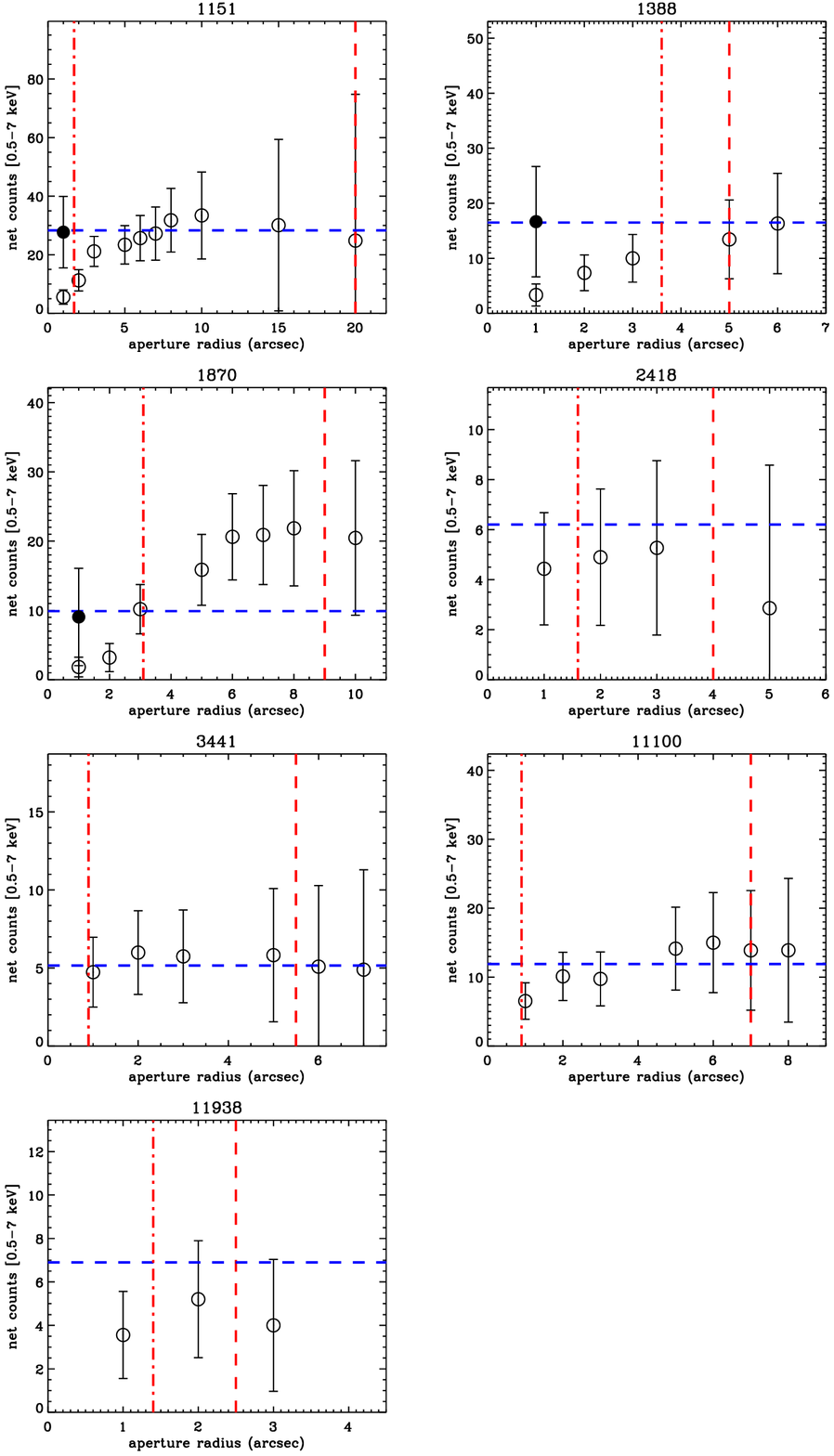}
  \caption{Aperture photometry for each off-nuclear candidate. The vertical dot-dashed line indicates the distance between the X-ray position and the centroid of the host galaxy; the vertical dashed line is the Petrosian radius of the host galaxy; the horizontal line corresponds to the counts estimated by EMLdetect \citep{puccetti09}. For objects XID$=$ 1151, 1388, and 1870, the filled circles represent the photometry on an area including 90$\%$ of the PSF obtained applying an aperture correction factor to the photometry measured on an aperture of 1\arcsec radius (see discussion in Sec. \ref{sec: sample}).}
              \label{aper_phot}%
    \end{figure*}

\section{Sample selection}
\label{sec: sample}

We have selected off-nuclear X-ray candidates from the Chandra COSMOS
Survey (C-COSMOS), which is a recently completed 1.8 Ms Chandra
program to image the central 0.9 deg$^2$ of the COSMOS field with an
effective exposure ranging from $\sim160$ ksec to $\sim80$ ksec going
from the center to the borders of the field (\citealt{elvis09}). The
limiting source detection depths are $1.9\times 10^{-16}$ erg
cm$^{-2}$ s$^{-1}$ in the [0.5-2 keV] band, $7.3\times 10^{-16}$ erg
cm$^{-2}$ s$^{-1}$ in the [2-10 keV] band, and $5.7\times10^{-16}$ erg
cm$^{-2}$ s$^{-1}$ in the [0.5-10 keV] band. We used a point source
catalogue including 1761 objects detected in at least one band (0.5-2,
2-7 and 0.5-7 keV) with a maximum likelihood ratio larger than
detml$=10.8$, corresponding to a probability of $\sim 2\times 10^{-5}$
that a catalog source is instead a background fluctuation
(\citealt{puccetti09}). The optical and infrared identifications of
almost all ($99.7\%$) of the sources are reported in
\citet{civano09}\footnote{The ULX candidates presented in this paper
  are flagged as 'off-nuclear' sources in \citet{civano09}.}. 

 As a first step to select off-nuclear X-ray sources, we verified the
 X-ray position accuracy that we have in the C-COSMOS observations
 following the procedure presented in Sec. 4.3 of
 \citet{puccetti09}. A set of 49 Chandra ACIS-I pointings has been
 simulated with the MARX\footnote{http://space.mit.edu/CXC/MARX.}
 simulator, adopting the same exposure times, aim points, and
 roll-angles as the real C-COSMOS pointings. The detection code
 PWDetect \citep{damiani97} was applied to the simulated data. We then
 compared the output of the detection algorithm with the input
 catalogue of the simulation. In Fig. \ref{pos-acc}, we show the
 distribution of the difference between the detection algorithm
 positions and the input positions in units of the X-ray positional
 error. The latest has been estimated as the ratio of the PSF at the
 position of the source and the square root of the net, background
 subtracted, source counts. In comparison with Fig. 10 of
 \citet{puccetti09}, we have restricted the analysis only to sources
 that have been detected at least in one image at an off-axis angle
 smaller than 5\arcmin~ to take advantage of an excellent PSF. From
 the distribution in Fig. \ref{pos-acc}, we find that $94\%$ of the
 sources have offsets below 1.8 times the positional error. We will
 adopt this value as a threshold to select off-nuclear candidates and
 therefore we expect that up to $6\%$ of our sample is contaminated by
 nuclear X-ray sources with large astrometric errors. We will shortly
 come back to this issue. Another possible source of spurious
 off-nuclear objects could be a poor astrometric accuracy of the X-ray
 images. According to Fig. 6 of \citet{elvis09}, $95\%$ of the Chandra
 sources have an absolute astrometric accuracy better than
 1.4\arcsec. For our study we aim at even better astrometric accuracy,
 therefore we have considered only the X-ray sources with an X-ray
 positional error smaller that 0.8\arcsec. We show the comparison
 between X-ray coordinates and optical coordinates for sources with a
 secure identification in Fig. \ref{ast-acc}: $95\%$ of the X-ray
 sources have an absolute astrometric accuracy better than 0.9\arcsec.

Summarizing, the off-nuclear candidates were selected using the following
criteria:
\begin{itemize}

\item[a)] The distance between the X-ray centroid and the optical centroid
  has to be larger than 1.8 times the radius of the Chandra positional
  error circle at that position.

\item[b)] The X-ray positional error is smaller than 0.8\arcsec.

\item[c)] The source has been observed in at least one Chandra
  pointing at an off-axis angle smaller than 5\arcmin.

\item[d)] The optical counterpart is a bright galaxy (R$_{\rm
AB}<$22).

\item[e)] The redshift of the host galaxy is less than z=0.3. The
  projected linear distance corresponding to an average Chandra
  positional error is $\sim4$ kpc at z=0.3. This means we will
  consider only off-nuclear candidates that are more than $\sim7$ kpc
  away from the center of the galaxy at z=0.3. At larger redshifts we
  would be able to select only off-nuclear candidates that are at
  larger distances ($> 7$ kpc) from the host galaxy center, where the
  number of observed off-nuclear sources seems to decrease
  (\citealt{swartz04}) and we would be more affected by the
  contamination of background objects. Therefore we limit our sample
  to z$<0.3$.

\item[f)] The distance between the X-ray centroid and the optical centroid
  is larger than 0.9\arcsec and smaller than the Petrosian radius
  (\citealt{petrosian76}, R$_{\rm P}$\footnote{R$_{\rm P}$ is defined
    as the radius at which the ratio (r$_{\rm P}$) of the local
    surface brightness at that radius and the mean surface brightness
    within that radius equals some specified value r$_{\rm
      P,lim}$. For a surface brightness distribution described by a de
    Vaucouleurs or an exponential profile, a value r$_{\rm P,lim}=0.2$
    is reached at R$_{\rm P}\sim 1.8$ R$_{\rm 1/2}$ and R$_{\rm P}\sim
    2.2$ R$_{\rm 1/2}$, respectively (R$_{\rm 1/2}$ is the half-light
    radius of the galaxy, see Fig. 17 of \citealt{scarlata07}).  }) of
  the galaxy, which we use as a measure of the galaxy's extension.

\end{itemize}

 If we consider the selection criteria b), c), d), and e), only
 sixteen sources from the C-COSMOS catalogue satisfy all of
 them. Based on our previous discussion of Fig. \ref{pos-acc} we
 expect up to $6\%$ of our sample to be due to nuclear X-ray sources
 with large astrometric errors, therefore we can conclude that our
 sample of ULXs contains less than one misidentified AGN.

For all the candidates provided by these selection criteria we have
verified that no other counterpart closer to the Chandra position was
present in any band from the $u^{\ast}$ ($\lambda_{center}=374.3$ nm)
filter to 24 micron. After this one-by-one check, we were left with
seven off-nuclear source candidates. Cutouts of these objects,
obtained from the COSMOS HST/ACS F814W imaging
(\citealt{koekemoer07}), are shown in Fig. \ref{mosaic1}, together
with the corresponding Chandra [0.5-7 keV] image.

 Each one of our off-nuclear sources has an estimate of the X-ray flux
 in the [0.5-7] keV band reported in \citet{elvis09}. These fluxes are
 derived from the counts estimated by
 EMLdetect\footnote{http://xmm.esac.esa.int/sas/8.0.0/emldetect},
 corrected to an area including $90\%$ of the PSF
 \citep{puccetti09}. In some cases such an area is large enough to
 include the whole host galaxy and therefore the X-ray flux could be
 the total integrated flux of the host galaxy itself. This would
 include the contribution from the population of X-ray binaries in the
 host, emission from diffuse gas and a possible weak central AGN. In
 order to estimate these possible contaminations on the measured X-ray
 fluxes, we have performed aperture photometry for each off-nuclear
 source. The radii of the apertures have been chosen with increasing
 size from a minimum of 1\arcsec~ up to include the whole galaxy. In
 Fig. \ref{aper_phot}, we plot the net counts in the [0.5-7] keV band
 as a function of the aperture radius. For four of our sources (XID=
 2418, 3441, 11100, 11938) the counts measured at different apertures
 are constant within the uncertainties. Therefore, we assume that the
 contribution of the host galaxy is not significant compared with the
 uncertainties on the measure. For the remaining three sources (XID=
 1151, 1388, 1870) the counts rise with the aperture radius and there
 may be a significant contamination due to the integrated flux of the
 whole galaxy. In order to minimize this contamination, we have
 considered the measured counts in the smaller aperture (1\arcsec
 ). We have then used the known PSF shape at the position of the
 source to estimate the expected fraction between the counts measured
 in an aperture of 1\arcsec and the ones over an area corresponding to
 $90\%$ of the PSF. We have then used this ratio to convert our
 measured counts on the 1\arcsec aperture into the expected ones on a
 $90\%$ PSF area. These corrected counts are indicated with a filled
 circle in the plots of Fig. \ref{aper_phot}, and we have used them in
 order to estimate the X-ray fluxes.\\ Full band 0.5-7 keV fluxes and
 errors were computed converting counts rates to fluxes using the
 formula: Flux=B$_{rate}$/(CF$\star 10^{11}$), where B$_{rate}$ is the
 count rate estimated as described above, and CF is the energy
 conversion factor. This conversion factor varies with the energy band
 and the spectral index $\Gamma$ assumed for the power-law
 spectrum. We have used the correction factor CF$=0.89$ counts
 erg$^{-1}$ cm$^2$ reported in Tab. 4 of \citet{elvis09} obtained for
 the 0.5-7 keV band and $\Gamma=1.7$. We decided for this average
 value of the spectral index following the study of \citet{swartz04}
 that has found a mean power-law index of $\Gamma=1.74\pm0.03$ for a
 sample of 154 ULX candidates observed with Chandra. We finally report
 on Tab. \ref{table:1} the 0.5-7 keV luminosities and errors for the
 seven off-nuclear sources. All sources have luminosities well in
 excess of $10^{39}$ erg s$^{-1}$ in the 0.5-7 keV band (the lowest
 X-ray luminosity in this band is $\approx 9 \times 10^{39}$ erg
 s$^{-1}$) and are therefore classified as ULX sources, using either
 spectroscopic or photometric redshifts.

   \begin{figure}
   \centering
   \includegraphics[width=8cm]{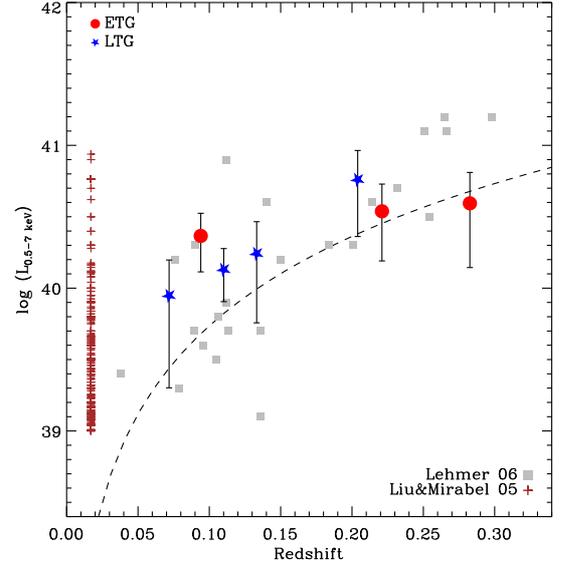}
      \caption{X-ray luminosity in the 0.5-7 keV band vs. redshift of the seven off-nuclear sources. The different symbols correspond to the host galaxy classification based on morphology and rest-frame colours: circles are ETGs, while stars are LTGs (see Sec. \ref{section:classification}). Squares are the off –nuclear sources from \citet{lehmer06}; crosses are the collection of local off-nuclear sources by \citet{liu05}.  The dashed line corresponds to the flux limit in the deepest region of the C-COSMOS survey: S$_{\rm lim}[0.5-7]=4.7\times 10^{-16}$ cgs.
              }
         \label{lx_z}
   \end{figure}

%


We have secure spectroscopic redshifts for four host galaxies from
zCOSMOS VIMOS observations at VLT (\citealt{lilly07,lilly09}). For the
remaining three objects we have used the extremely accurate
photometric redshifts available in the COSMOS field
(\citealt{ilbert08,salvato08}) based on 30 broad, intermediate, and
narrow bands from the UV to the mid-IR. We show in Fig. \ref{lx_z} the
X-ray luminosity in the [0.5-7] keV band versus redshift of the seven
ULXs. The X-ray luminosities were computed according to the formula:

\begin{eqnarray}
L_X=4\pi d^2_Lf_X(1+z)^{\Gamma-2}
\end{eqnarray}

where $d_L$ is the luminosity distance, $f_X$ is the X-ray flux in the
[0.5-7] keV band, and $\Gamma$ is the X-ray photon index. We assumed
$\Gamma=1.7$ (see discussion in this section). Different symbols
correspond to the morphological classes of the host galaxies (see Sec.
\ref{section:classification}). Squares are the off-nuclear sources
from \citet{lehmer06}; crosses are the collection of local ULXs by
\citet{liu05}.  The dashed line corresponds to the flux limit of the
C-COSMOS survey, S$_{\rm lim}[0.5-7]=4.7\times 10^{-16}$ erg cm$^{-2}$
s$^{-1}$.

In order to estimate how many background sources we expect to
contaminate our sample, we applied a random shift between 30\arcsec
and 2\arcmin~ to the C-COSMOS sources and searched for chance
coincidences with R$_{\rm AB}<$22 and z$<0.3$ galaxies. We repeated
this procedure $10,000$ times and we found that, on average, the
chance coincidences are $\approx 1.3$. Only for $2\%$ of the $10,000$
simulations we found more than 3 chance coincidences. Summarizing, we
expect less than one misidentified AGN due to astrometric errors and
on average 1.3 serendipitous background source matches.

   \begin{figure}
   \centering \includegraphics[width=8cm]{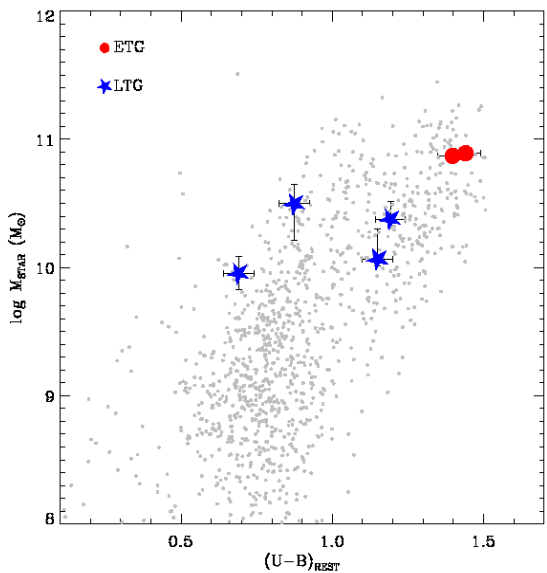}
      \caption{Colour-mass diagram: circles and stars are respectively
        ULX host galaxies classified as ETGs and LTGs based on their
        morphology/colours; the dots are galaxies in the C-COSMOS area
        with z$<0.3$ and R$_{\rm AB}<22$. For the source XID=1151 the
        photometric coverage is limited to few bands and we cannot
        constrain its stellar mass.}
         \label{UB_mstar}
   \end{figure}

\section{Host galaxy properties}

\subsection{Galaxy classification} 
\label{section:classification}

Studies of local samples of ULXs (e.g. \citealt{swartz04}) have shown
that these sources are mainly present in late type galaxies. A visual
inspection of Fig. \ref{mosaic1} suggests that the ULXs at
intermediate redshifts that we are studying are hosted in both early
and late type galaxies (ETGs and LTGs, hereafter). 

To confirm this impression we classified the host galaxies based on
their morphology and colours (e.g. \citealt{mignoli09}). Taking
advantage of the COSMOS HST/ACS F814W images (\citealt{koekemoer07}),
we have used an accurate morphological classification derived by
\cite{scarlata07} through the Zurich Estimator of Structural Type
(ZEST). \cite{scarlata07} describe in detail the methodology and the
performances of this method. We only recall here that the ZEST
classification is based on: a) five non-parametric diagnostics
(asymmetry A, concentration C, Gini coefficient G, 2$^{nd}$ order
moment of the brightest $20\%$ of galaxy pixels M$_{20}$, ellipticity
$\epsilon$); and b) the exponent $n$ of single Sersic fits to the
two-dimensional surface brightness distributions. ZEST assigns to each
galaxy a morphological type (1=early type; 2=disk; 3=irregular) and a
bulgeness parameter that splits the disk galaxies in four separate
bins, from bulge dominated disks (2.0) to pure disk galaxies
(2.3). For the bulge-dominated galaxies (2.0), we complemented the
morphological information with their rest-frame colours to further
subdivide them: if they have red U-B rest-frame colours we include
them in the ETGs sample (XID$= 1151$), otherwise we classify them as
LTGs (XID$= 1870$). In Fig. \ref{UB_mstar} we plot the colour-mass
diagram for our ULX host galaxies: they can be divided into three ETGs
and four LTGs. We will describe in Sec. \ref{sfr_sm} the method used
to estimate stellar masses.

The slight preference for ULXs to be hosted in LTGs could be explained
by the different shapes of the X-ray luminosity function (XLF) for
Low-Mass X-ray Binaries (LMXBs) and High-Mass X-ray Binaries (HMXBs)
derived for local galaxies (\citealt{grimm03,gilfanov04}): the former
has an abrupt cut-off at L$_{\rm X}\approx 10^{39}$ erg s$^{-1}$,
while the latter can be described with a power-law with a slope
$\alpha = 1/6$. Since early-type stars are the dominant stellar
population of LTGs, we expect X-ray binaries with O or B type
companions, HMXBs, to be common in these objects. This translates into
a higher chance to detect ULXs in LTGs or, in any case, in galaxies
with current star formation activity.

\subsection{Stellar masses and SFR}
\label{sfr_sm}

Stellar masses (M$_\star$) are derived from the stellar population
synthesis model that represents the best fit to the observed
photometry (from the $u^{\ast}$ band to 4.5 $\mu$m) using a $\chi ^2$
minimization technique. The procedure is explained in detail by
\citet{bolzonella09}. Here we only recall the basic ingredients of the
Spectral Energy Distribution (SED) fitting procedure:
\begin{itemize}
\item stellar population synthesis models from the libraries of \citet{bruzual03};
\item eleven ``smooth'' star formation histories for each library: one constant star formation model plus 10 $\tau$-model with e-folding time-scales $\tau = 0.1, 0.3, 0.6, 1, 2, 3, 5, 10, 15, 30$ Gyr;
\item a Chabrier initial mass function;
\item a Calzetti extinction law with  $0<A_{\rm V}<3$;
\item solar metalicity (Z$=$Z$_\odot$).
\end{itemize}

   \begin{figure}
   \centering
   \includegraphics[width=8cm]{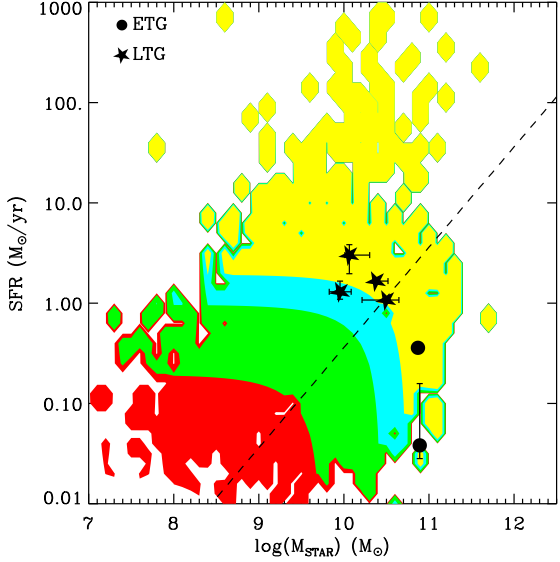}
      \caption{SFR versus stellar masses of the galaxies in the comparison sample (see text). The contours corresponds to region where more than 0.1 (red), 1 (green), 5 (cyan), and 10 (yellow) X-ray off-nuclear sources per galaxy are expected. The symbols show the location in this plane of the host galaxies of the ULXs. For the source XID=1151 the photometric coverage is limited to few bands and we cannot constrain its stellar mass. Symbols are the same of Fig. \ref{lx_z}. The dashed line is the locus where we expect the same number of LMXBs and HMXBs with L$_{\rm X}>10^{38}$ erg s$^{-1}$. Above this line the number of HMXBs is expected to be higher than that of LMXBs.}
         \label{mstar_sfr}
   \end{figure}

The star formation rate (SFR) values have been estimated using when
possible (for three ULXs) the H$\alpha$ $\lambda$6563 line flux as
measured by the routine Platefit (\citealt{lamareille09}), with a
correction for reddening. We used the \citet{kennicutt98} relation
between H$\alpha$ and SFR: SFR(M$_\odot$ yr$^{-1})=(7.9\times
10^{-42})$L($H\alpha$) ergs s$^{-1}$. The de-reddened flux of
H$\alpha$ has been computed according to the formula: $F_{der}=F_{obs}
\times 10^{c[1+f(\lambda)]}$ where $f(\lambda)=3.15854 \times
10^{-1.02109\lambda}-1$ and $c=1.47E_{B-V}$
(\citealt{seaton79,maier05}). If a measure of the H$\beta$ flux was
available we have estimated E$_{B-V}$ from the Balmer decrement,
adopting the \citet{odonnell94} Milky Way extinction curve. Otherwise,
we used the average value $<E_{B-V}>\sim 0.2$ mag derived by
\citet{moustakas06}.  For the sources without H$\alpha$ in the
spectral range or for which we have only a photometric redshift, we
have used the SFR estimate from the SED fitting procedure.

From the COSMOS catalogue (\citealt{capak07,ilbert08}), we have
selected a comparison sample of galaxies inside the area covered by
Chandra. We imposed the same constraints used to select the
off-nuclear candidates: z$<$0.3 and R$_{\rm AB}<22$. We have also
removed all the sources that are best fitted by stellar SED templates
(\citealt{ilbert08}). At the end, the comparison sample consists of
2066 galaxies. For all of them we have derived stellar masses and SFR
values as described above.

We now estimate the probability to have an off-nuclear source given a
host galaxy with a particular stellar mass and SFR. We will consider
both LMXBs and HMXBs.

For LMXBs we used the average XLF derived by \citet{gilfanov04}. This
is described by a powerlaw with two breaks, from their formula (8):

\begin{eqnarray}
\frac{dN}{dL_{38}}=\left\{ \begin{array}{ll}
\renewcommand{\arraystretch}{3}
K_1 \left(L_{38}/L_{b,1}\right) ^{-\alpha_1}	
			& \mbox{\hspace{0.9cm} $L_{38}<L_{b,1}$}\\
K_2 \left(L_{38}/L_{b,2}\right)^{-\alpha_2}	
			& \mbox{$L_{b,1}<L_{38}<L_{b,2}$}\\
K_3 \left(L_{38}/L_{cut}\right)^{-\alpha_3}	
			& \mbox{$L_{b,2}<L_{38}<L_{cut}$}\\
0			& \mbox{\hspace{0.9cm} $L_{38}>L_{cut}$}\\
\end{array}
\right.
\label{eq:uxlf}
\end{eqnarray}

where $L_{38}=L_X/10^{38}$ erg/s and normalizations $K_{1,2,3}$ are defined as:

\begin{eqnarray}
K_2=K_1 \left(L_{b,1}/L_{b,2}\right)^{\alpha_2}\nonumber \\
K_3=K_2 \left(L_{b,2}/L_{cut}\right)^{\alpha_3}\nonumber 
\end{eqnarray}

We used the best fitting parameter derived by \citet{gilfanov04}:
$\alpha_{1}=1.0$, L$_{b,1}=0.19$, $\alpha_{2}=1.86$, L$_{b,2}=5.0$,
$\alpha_3=4.8$. The high-luminosity cut-off was fixed at
L$_{cut}=500$. For the average normalization we used the best fitting
value given by \citet{gilfanov04}, $K_1=440.4\pm25.9 {\rm ~ per~
  10^{11}~ M_\odot}$, and we will assume a linear relation between the
number of X-ray sources and stellar mass as found by the same authors
(see Sec. 5 of \citealt{gilfanov04}).  We note that up to L$_{\rm
  X}\approx 2\times 10^{39}$ ergs s$^{-1}$ the XLF of
\citet{gilfanov04} is consistent with later studies (e.g. see Fig. 14
of \citealt{humphrey08}). Above this luminosity we are extrapolating
the XLF since no data are currently available and therefore the
uncertainties are large. For the slope at the highest luminosities, we
have considered values in the range $\alpha_3=[2,6]$ and we did not
find any significant difference from the contours reported in
Fig. \ref{mstar_sfr}.

For the HMXBs we used instead the luminosity function derived by
\citet{grimm03}. In particular, we used the cumulative form of it, 
corresponding to their formula (7):

\begin{eqnarray}
N(>L_{38})=5.4~SFR \left(L_{38}^{-0.61}-210^{-0.61}\right)\nonumber \\
\end{eqnarray}

where SFR is in units of $M_\odot$ yr$^{-1}$.

We have then calculated the number of X-ray binaries with L$_{\rm
  X}>10^{38}$ erg s$^{-1}$ that we expect in each galaxy integrating
the XLFs for a given SFR and M$_\star$. In Fig. \ref{mstar_sfr} we
show the contours corresponding to regions where we expect more than
0.1 (red), 1 (green), 5 (cyan), 10 (yellow) X-ray sources with L$_{\rm
  X}>10^{38}$ erg s$^{-1}$. In reality these numbers have to be
considered upper limits because we have not taken into account the
limited Chandra spatial resolution that does not allow to detect
off-nuclear sources with small offsets (see Fig. 10 of
\citealt{lehmer06}).

From Fig. \ref{mstar_sfr}, we find that all our ULX candidates are
hosted in galaxies for which a large number of X-ray binaries is
predicted. The dashed line in Fig. \ref{mstar_sfr} is the locus where
we expect the same number of LMXBs and HMXBs with L$_{\rm X}>10^{38}$
erg s$^{-1}$. This line clearly divides a region (below the line)
where the XLF of LMXBs is dominating and therefore the contours are
mainly defined by the M$_{\star}$ values, from a region (above the
line) where the HMXBs are more numerous and the contours are
determined by the level of the SFR. Our morphological classification
is consistent with this picture: ETGs, characterized by lower SFR and
high stellar masses, are located in the bottom-right part of the plot,
where the expected number of LMXBs is higher than the number of
HMXBs. However, we note that there are suggestions in the literature
that no ULX LMXBs may actually exist. ~\citet{irwin04} had shown that
the number of ULXs detected in a sample of 28 ellipticals observed
with Chandra is equal to the number of expected foreground/background
objects. Additionally, such ULXs are uniformly distributed and do not
follow the optical light of the galaxies. \citet{irwin04} also
verified that the same statements can be made for the ULXs associated
to early-type galaxies presented in \citet{colbert02}.

It would be interesting to repeat the same computation that generated
Fig. \ref{mstar_sfr} considering only X-ray binaries with L$_{\rm
  X}>10^{39}$ erg s$^{-1}$, and therefore to be able to verify the
hypothesis that ULXs are the high-luminosity tail of normal X-ray
binaries. Unfortunately, this is not possible due to the
poor-knowledge of the high luminosity slope of the XLF for LMXBs and
HMXBs.

   \begin{figure} \centering
   \includegraphics[width=8cm]{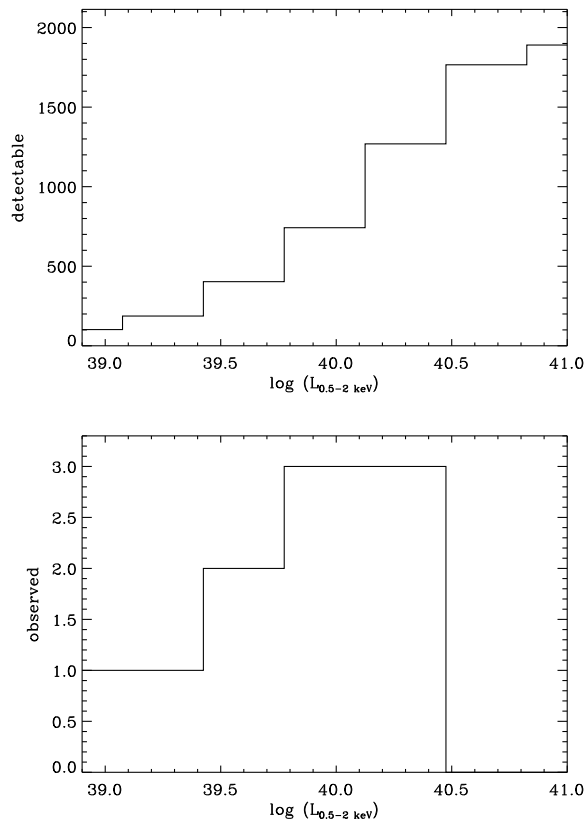} \caption{Top panel: the
   number of galaxies for which we could detect an off-nuclear source
   of a given 0.5-2 keV luminosity L$_{0.5-2~keV}$ or higher. Bottom panel: the observed number of galaxies in each L$_{0.5-2~keV}$ bin hosting an ULX of luminosity
L$_{0.5-2~keV}$ or higher.  } 
   \label{detectable} \end{figure}

   \begin{figure}
   \centering
   \includegraphics[width=8cm]{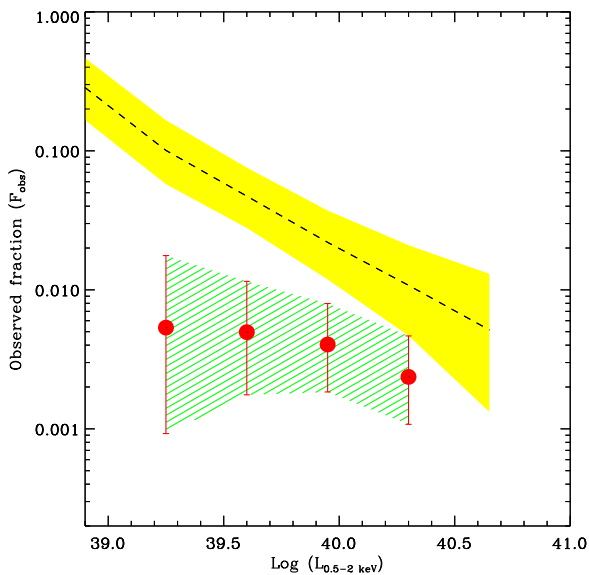}
      \caption{The observed fraction of galaxies with an off-nuclear source with a luminosty of L$_{0.5-2~keV}$ or greater. The red points and associated 1$\sigma$ confidence region are from our sample, while the dashed line and the 1$\sigma$ confidence region have been obtained by \citet{lehmer06} from the Chandra Deep Fields.              }
         \label{frac_off}
   \end{figure}

\section{Fraction of galaxies hosting a ULX}

As already pointed out by \citet{ptak04}, useful constraints on the
nature of ULXs can be obtained deriving the fraction of galaxies that
harbor a ULX as a function of the X-ray luminosity. For example,
\citet{kording02} have compared the luminosity distribution of X-ray
point sources in nearby galaxies with that predicted by X-ray
population synthesis models to check whether microblazars
(microquasars with relativistically beamed jets pointing towards the
observer) may represent an alternative to the intermediate mass black
holes scenario for ULXs. In order to compute this fraction, we have
used the comparison sample selected in Sec. \ref{sfr_sm}. We derived
for each individual galaxy a $90\%$ upper limit on its X-ray flux in
the [0.5-2] keV band according to the procedure described in Sec. 6.5
of \citet{puccetti09}, to which we refer the reader for details. The
top panel of Fig. \ref{detectable} shows the number of galaxies for
which we could detect an off-nuclear source of 0.5-2 keV luminosity
L$_{0.5-2~keV}$ or larger. The bottom panel of the same figure shows
the observed number of galaxies in each L$_{0.5-2~keV}$ bin hosting an
ULX of luminosity L$_{0.5-2~keV}$ or larger. In order to derive the
observed fraction of galaxies with an off-nuclear source, we divided
the values of the histogram in the bottom panel by those in the top
panel of Fig. \ref{detectable}. The result is shown in
Fig. \ref{frac_off}. The red points are the result of our analysis,
and the dashed area is the 1$\sigma$ confidence region computed using
the prescriptions for small numbers statistic by
\citet{gehrels86}. For comparison, we report in the same figure also
the fractions obtained by \citet{lehmer06} from the Chandra Deep
Fields (dashed line and 1$\sigma$ confidence region). These fractions
should be considered as lower limits due to the limited Chandra
spatial resolution that does not allow to detect off-nuclear sources
with small offsets (see Fig. 10 of \citealt{lehmer06}). The agreement
between our results and the CDFs points is reasonably good above
log($L_{0.5-2~keV})> 40$, although our point and the associated
confidence contours are about a factor of two lower than, but
consistent with, those derived by \citet{lehmer06}. In the lower
luminosity bins, it seems that the two measures are discrepant;
however, we do not consider this difference highly significant, since
the measured fractions are consistent at the 2$\sigma$ level. Also, at
the faintest fluxes the differences between the two X-ray catalogs
used is more severe. For these faint sources the positional
uncertainties affecting our sample are larger than for the same
sources detected in the longer Chandra exposures of the CDFs, and
therefore we may be missing the faintest ULXs in the sample if their
error box is consistent with the position of the nucleus. We also note
that our selection criteria for off-nuclear sources reported in
Sec. \ref{sec: sample} are more conservative than the ones used by
\citet{lehmer06}. From Fig. \ref{frac_off} we found that $\approx
0.5\%$ and $\approx0.2 \%$ of the galaxies are hosting a ULX with
L$_{0.5-2~keV}\gtrsim 3 \times 10^{39}$ and L$_{0.5-2~keV}\gtrsim 2
\times 10^{40}$ erg s$^{-1}$, respectively.

%
\begin{table*}
\begin{minipage}[t]{\columnwidth}
\caption{Properties of ULXs in C-COSMOS}             
\label{table:1}      
\centering       
\renewcommand{\footnoterule}{}  
\begin{tabular}{rccccccccc}     
\noalign{\smallskip}
\hline\hline   
\noalign{\smallskip}
XID\footnote{ID of the Chandra source (Elvis et al. 2009)} & RA & Dec & Counts\footnote{X-ray counts in the [0.5-7] keV band.} & log L$_{\rm X}$\footnote{Logarithm of the [0.5-7] keV X-ray luminosity.} & Pos. error\footnote{X-ray positional error.} & Offset & Offset  & Offset\footnote{Ratio of the distance between the X-ray centroid and the optical centroid over the radius of the Chandra positional error circle.} & Off-axis\footnote{Off-axis angle value in the image where the source is closer to the on-axis position.} \\
 & \multicolumn{2}{c}{(J2000)\footnote{X-ray coordinates of the ULX.}}  & (0.5-7 keV)  & (erg s$^{-1}$)& (arcsec) & (arcsec) & (kpc) & & (arcmin)  \\
\noalign{\smallskip}
\hline 
\noalign{\smallskip}

              1151    &         10:00:10.39    &         02:09:23.40    &                28    &        40.4$^{40.5}_{40.1}$    &         0.5    &        1.67    &        2.91    &        3.42  & 2.8   \\
              1388    &         10:01:08.46    &         02:01:06.05    &                17    &        40.8$^{40.9}_{40.4}$    &         0.6    &        3.62    &       12.12    &        5.73   & 2.6  \\
              1870    &         10:01:03.76    &         02:30:50.22    &                 9    &        39.9$^{40.2}_{39.3}$    &         0.4    &        3.12    &        4.66    &        7.25   & 2.6 \\
              2418    &         10:00:08.43    &         02:14:47.65    &                 6    &        40.6$^{40.8}_{40.1}$    &         0.3    &        1.58    &        6.76    &        4.53    & 1.3 \\
              3441    &         09:59:33.78    &         01:49:06.92    &                 5    &        40.2$^{40.5}_{39.8}$    &         0.5    &        0.95    &        2.26    &        1.87    & 3.7 \\
             11100    &         10:00:58.65    &         02:11:39.90    &                12           &        40.1$^{40.3}_{39.9}$    &         0.4    &        0.92    &        1.85    &        2.37  & 3.4  \\
             11938    &         10:00:43.02    &         02:00:32.74    &                 7           &        40.5$^{40.7}_{40.2}$    &         0.8    &        1.39    &        4.79    &        1.77  & 4.4  \\

\hline                  
\end{tabular}
\end{minipage}
\end{table*}
%

We now discuss the observed trend of the fraction of ULX as a function
of their X-ray luminosities in the frame of the beaming model of
\citet{king09}. According to this model, ULX are stellar mass black
holes accreting at a super-Eddington rate ($\dot m\equiv \dot M/\eta
L_{\rm Edd} c^2 > 1$, for a typical radiative efficiency $\eta\sim
0.1$ and accretion rate $\dot M$). Matter accreting at such rates is
easily blown away close to the inner edge of the accretion disc
(\citealt{shakura73}); then, the radiative output from the resulting
flow pattern is of the order of $L\approx L_{\rm Edd}(1+\ln \dot m)$,
but emerges collimated by the central funnel with a
beaming\footnote{Note that here 'beaming' simply means geometrical
  collimation, and not relativistic beaming.} factor $b\propto \dot
m^{-2}$, so that an external observer who happens to have its line of
sight within the beaming cone would infer a spherical luminosity:
$L_{\rm ULX}\simeq 10^{39} m_7 (1+\ln \dot m)/b$ erg/s (where $m_7$ is
the black hole mass in units of 7 solar masses; see \citealt{king09}
for further details). Thus, neglecting the weak logarithmic dependence
on $\dot m$, this model directly links the observed luminosity of a
ULX with its beaming factor $b$.

Let us now consider a population of ULX with host galaxy space density
(as a function of distance $d$): $n_g(d)$ Mpc$^{-3}$. The results of
\citet{lehmer06} imply an almost linear decline of the cumulative
number of ULX per galaxy with observed luminosity, $F_{\rm obs}\simeq
F_0 (L_{\rm ULX}/10^{39})^{-1}$, where $F_0\simeq 0.1$ is the observed
fraction of galaxies hosting a ULX with $L_{\rm ULX} > 10^{39}$. The
differential fraction $\Phi_{\rm obs}$, i.e. the fraction of galaxies
containing a ULX with luminosity $L_{\rm ULX}$ per unit logarithmic
interval of luminosity can be derived by simply differentiating the
above expression, to obtain $\Phi_{\rm obs}\equiv dN/dLogL_{\rm
  ULX}=F_0 (L_{\rm ULX}/10^{39})^{-1}\approx F_0 b/m_7$, where the
last approximate equality has been derived neglecting the logarithmic
dependence of $L_{\rm ULX}$ on $b$.

We now consider the application of such a model to a multi-wavelength
survey like COSMOS. We define the limiting flux of the survey in the
X-ray band as $f_{\rm lim}=f_{\rm -16} \times 10^{-16}$ erg s$^{-1}$
cm$^{-2}$, so that an object of beaming factor $b$ can be seen out to
a distance of $d(b)=(10^{39}m_7/4\pi f_{\rm lim}b)^{1/2}\simeq 313
(b/m_7)^{-1/2} f_{\rm -16}^{-1/2}$ Mpc, and express, in full
generality, the number density of galaxies as a function of distance
as $n_g(d)=n_g(d(b))\equiv n_{g,0}(b/m_7)^{\alpha}$, where $n_{g,0}$
is the number density of possible host galaxies in the survey at the
maximum distance where an un-beamed source ($b=1$) can be seen. Such
an expression is a very general form appropriate for power-law
luminosity functions in Euclidean Universes and is adopted here for
the sake of simplicity\footnote{Although we have applied a
  k-correction to the luminosity values in equation (1), we resolved
  to make the calculations in this paragraph under the assumption of a
  Euclidean Universe to simplify the derivation of equation (5).}; the
exponent $\alpha$ depends both on the galaxy luminosity function slope
and on the survey selection function and can in principle be derived
empirically for any given survey: typically, for flux-limited surveys,
we have $\alpha>0$, while volume limited ones have $\alpha\approx 0$.
Given the observed cumulative fraction $F_{\rm obs}$\footnote{We
  assume in this calculation that the fraction of galaxies hosting a
  ULX does not change as a function of distance. This is an
  approximation, since the star formation rate varies with redshift
  and therefore it is plausible that the fraction of galaxies hosting
  a ULX varies too.}, and the corresponding differential $\Phi_{\rm
  obs}=F_0b/m_7$ one has to search through a space volume $V\sim
1/n_g(d)F_0(b/m_7)$ to find a ULX with beaming factor $b$ (within a
unit logarithmic interval of b). From this expression for the volume
we derive:
\begin{equation}
d(b)=125\left(A \frac{n_{g,0}}{0.05}\frac{F_0}{0.1}\right)^{-1/3}\left(\frac{b}{m}\right)^{-(1+\alpha)/3}
\end{equation}
Thus, the minimum beaming factor (corresponding to the maximal
luminosity) of a ULX in a survey of area $A$ (in units of square
degrees), is given by:
\begin{equation}
b_{\rm min}=0.4^{\gamma}\left(f_{\rm
    -16}^{\gamma/2}A^{-\gamma/3}\right)
\left(\frac{n_{g,0}}{0.05}\frac{F_0}{0.1}\right)^{-\gamma/3}m_7 
\end{equation}
where $\gamma=\frac{6}{2\alpha-1}$. The overall efficiency of finding
ULX scales as $\ln L_{\rm ULX,max}$. We recall that L$_{\rm ULX}$ is
the spherical luminosity that would be inferred by an external
observer who happens to have its line of sight within the beaming
cone. Applying this rough estimate with $\alpha=1$, m$_7=1$,
n$_{g,0}=0.05$, F$_0=0.1$ to the COSMOS survey ($f_{\rm -16}\simeq 2$,
$A=0.9$), we obtain $b_{\rm min}\simeq 0.04$, $L_{\rm ULX,max}\simeq
2.5 \times 10^{40}$ in reasonable agreement with the present
data. Interestingly, this also suggests that larger, but shallower,
surveys could be more efficient in finding ULX (provided a similarly
deep sample of host galaxies can be identified): the all sky eROSITA
survey ($f_{\rm -16}=100$, $A=4\times 10^4$) could find a large number
of ULX, including microblazars up to $b_{\rm min}=2.6\times 10^{-6}$,
$L_{\rm ULX,max}\simeq 4.0 \times 10^{44}$.

\section{Conclusions}

We have presented a sample of ultraluminous X-ray sources (ULXs)
selected from the Chandra survey in the COSMOS area (C-COSMOS). From
1761 X-ray sources detected with a maximum likelihood threshold of
detml=10.8 in at least one detection band, we have selected 7 ULX
candidates covering the redshift range z=0.072-0.283.

Taking advantage of the excellent ancillary data available in the
COSMOS field, we have studied the properties of their host
galaxies. From a detailed morphological analysis of the ACS images and
rest-frame colours, we found that ULXs are hosted both in late and in
early type galaxies, with a slight preference for the former.

From the multi-band photometry and from the optical spectral lines, we
have measured stellar masses and star formation rates for the host
galaxies. Using literature X-ray luminosity functions for HMXBs and
LMXBs, we have defined probability areas for having detectable
off-nuclear sources in the plane SFR versus M$_\star$. All our ULXs
candidates are hosted in galaxies for which we expect a large number
of X-ray binaries with L$_{\rm}>10^{38}$ erg s$^{-1}$.

The presence of IMBHs ($\sim 10^2-10^5 M_\odot$) in some of our ULXs
cannot be excluded with the current data. The best candidates for this
new class of accreting black holes are the ULXs hosted in early type
galaxies (therefore not associated with recent star formation
activity) and with X-ray luminosity above $10^{41}$ erg s$^{-1}$ that
can be difficult to explain with high-mass stellar black holes. The
objects that satisfy these criteria from our sample are XID$=$ 2418
and 11938. Longer X-ray exposures could give us more insights on the
real nature of these sources from a detailed study of the X-ray
spectrum. Similarly, we cannot set constraints on the recoiling
black-hole nature of our sources with the current data, but it is
worth mentioning that recent predictions by \citet{volonteri08} expect
at most one of such objects in the C-COSMOS survey, assuming the most
favorable scenario (spinning black holes, no bulge in the host galaxy,
long active phase).

Finally, we have derived the fraction of galaxies hosting a ULX as a
function of the X-ray luminosity. We found that $\approx 0.5\%$ and
$\approx0.2 \%$ of the galaxies are hosting a ULX with
L$_{0.5-2~keV}\gtrsim3 \times 10^{39}$ and L$_{0.5-2~keV} \gtrsim 2
\times 10^{40}$ erg s$^{-1}$. This is in reasonably good agreement
with the observed fraction derived in the Chandra Deep Fields by
\citet{lehmer06} above log(L$_{0.5-2 keV}>40$ erg s$^{-1}$. A possible
discrepancy in the lower luminosity bins can be likely attributed to
the differences in the limiting fluxes of the two catalogs and,
therefore, to the different positional uncertainties affecting faint
X-ray sources.

\begin{table}
\begin{minipage}[t]{\columnwidth}
\caption{Properties of the host galaxies of ULXs in C-COSMOS}             
\label{table:2}      
\centering       
\renewcommand{\footnoterule}{}  
\begin{tabular}{rcccccc}     
\noalign{\smallskip}
\hline\hline   
\noalign{\smallskip}
XID\footnote{ID of the Chandra source (Elvis et al. 2009)} & i mag & R$_{\rm P}$\footnote{Petrosian radius of the host galaxy \cite{petrosian76}} & z\footnote{Redshift of the host galaxy: ``s'' for spectroscopic and ``p'' for photometric redshifts.} &  Class\footnote{Morphological classification of the ULX host galaxy: early type galaxy (ETG) or late type galaxy (LTG).} & log(M$_\star$) & SFR \\
 & (AB) & (arcsec)  & & & (M$_\odot$) & (M$_\odot$/yr)\\
\noalign{\smallskip}
\hline 
\noalign{\smallskip}

              1151\footnote{The photometric coverage is limited to few bands and we cannot constrain its M$_\star$ or SFR}    &       15.62    &       19.89    &       0.094        $^{\rm p}$ &        ETG & ... & ... \\
              1388    &       20.39    &        5.20    &       0.204        $^{\rm p}$ &         LTG & 10.1$^{+0.2}_{-0.1}$ & 3.0$^{+0.8}_{-1.1}$ \\
              1870    &       18.40    &        9.21    &       0.072        $^{\rm s}$ &         LTG & 9.9$^{+0.2}_{-0.1}$ & 1.3$^{+0.4}_{-0.2}$\\
              2418    &       19.45    &        3.99    &       0.283        $^{\rm s}$ &         ETG & 10.9$^{+0.1}_{-0.1}$ & 0.3$^{+0.1}_{-0.1}$\\
              3441    &       18.19    &        5.50    &       0.133        $^{\rm s}$ &         LTG  & 10.5$^{+0.1}_{-0.3}$ & 1.1$^{+0.1}_{-0.1}$\\
             11100    &       18.48    &        6.90    &       0.110        $^{\rm s}$ &         LTG & 10.4$^{+0.1}_{-0.1}$ & 1.7$^{+0.1}_{-0.1}$\\
             11938    &       18.94    &        2.53    &       0.221        $^{\rm p}$ &         ETG & 10.9$^{+0.1}_{-0.1}$ & 0.04$^{+0.12}_{0.01}$\\

\hline                  
\end{tabular}
\end{minipage}
\end{table}

\begin{acknowledgements}
	This work is based on observations made with ESO Telescopes at
        the La Silla or Paranal Observatories under programme ID
        175.A-0839. We are grateful to the referee for detailed and
        extremely useful comments that improved the quality of the
        paper. We thank Piero Rosati and Bret Lehmer for useful
        scientific discussions. We are grateful to Bret Lehmer for
        providing the data points of the CDFs used in
        Fig. \ref{frac_off}. This work has been supported in part by
        the grants: ASI/COFIS/WP3110 I/026/07/0, ASI/INAF I/023/05/0,
        ASI I/088/06/0, PRIN/MIUR 2006-02-5203.
\end{acknowledgements}

\end{document}